\documentclass[10pt]{article}
 
\renewcommand{\title}[1]{\null\vspace{25mm}\noindent{\Large{\bf #1}}\vspace{10mm}}
\newcommand{\authors}[1]{\noindent{\large #1}\vspace{20mm}}
\newcommand{\address}[1]{{\center{\noindent #1\vspace{10mm}}}}
\renewcommand{\abstract}[1]{\vspace{17mm}
\noindent{\small{\em Abstract.} #1}\vspace{2mm}}     
 
\newcommand{\mnr}{{\mu\nu\rho}}
\newcommand{\mn}{{\mu\nu}}

\newcommand{\nm}{\nonumber}

\newcommand{\be}{\begin{equation}}
\newcommand{\ee}{\end{equation}}
\newcommand{\ba}{\begin{array}}
\newcommand{\ea}{\end{array}}
\newcommand{\bea}{\begin{eqnarray}}
\newcommand{\eea}{\end{eqnarray}}
\newcommand{\Li}{{\cal L}_\tau}
\newcommand{\del}{\delta_\tau}
\newcommand{\id}{i_\tau}
\newcommand{\bh}{\frac{1}{2}}
\newcommand{\bzd}{\frac{2}{3}}

\newcommand{\tr}{{\rm tr}\;}
\newcommand{\de}{\partial}
\newcommand{\intM}[1]{\int_{{\cal M}_#1}\tr}
\newcommand{\intMohnetr}[1]{\int_{{\cal M}_#1}}
\newcommand{\g}[1]{\Gamma^{(#1)}}
\newcommand{\V}[2]{\frac{\delta #1}{\delta #2}}
\newcommand{\ST}[2]{\V{\g{0}}{#1}\V{\g{0}}{#2}}
\newcommand{\LST}[2]{\V{\g{0}}{#1}\V{}{#2}+\V{\g{0}}{#2}\V{}{#1}}

\newcounter{saveeqn}
\newcommand{\alpheqn}{
        \setcounter{saveeqn}{\value{equation}}
        \stepcounter{saveeqn}
        \setcounter{equation}{0}
        \renewcommand{\theequation}{\mbox{\arabic{saveeqn}-\alph{equation}}}
        }
\newcommand{\reseteqn}{
        \setcounter{equation}{\value{saveeqn}}
        \renewcommand{\theequation}{\arabic{equation}}
        }

\begin{document}   \setcounter{table}{0}
 
\begin{titlepage}
\begin{center}
\hspace*{\fill}{{\normalsize \begin{tabular}{l}
                              {\sf hep-th/0002234}\\
                              {\sf REF. TUW 00-08}\\
			      {\sf \today}
                              \end{tabular}   }}

\title{A generalized $p$-form model in $D=3$}

\authors {T.~Pisar$^{1}$, J.~Rant$^{2}$, H.~Zerrouki$^{3}$}    \vspace{-20mm}
       
\address{Institut f\"ur Theoretische Physik,Technische Universit\"at Wien\\
      Wiedner Hauptstra\ss e 8-10, A-1040 Wien, Austria}
\footnotetext[1]{Work supported by the "Fonds zur F\"orderung der Wissenschaflicher Forschung", under Project Grant Number P11582-PHY.}
\footnotetext[2]{email: jrant@tph.tuwien.ac.at}
\footnotetext[3]{Work supported by the "Fonds zur F\"orderung der Wissenschaflicher Forschung", under Project Grant Number P13125-PHY.}

\end{center} 
\thispagestyle{empty}

\abstract{A topological model in three dimensions is proposed. It combines the Chern-Simons action with a BFK-model which was investigated recently by the authors of \cite{DelCima:1999kb}. The finiteness of the model to all orders of perturbation theory is shown in the framework of algebraic renormalization procedure.}

\end{titlepage}

\section{Introduction}

During the last decade the topological field theories (TFT) \cite{Birmingham:1991ty} have been an arena of large investigations. The TFTs are characterized by the fact that the observables depend only on the global structure of the space-time manifold on which the model is defined. In particular, this implies that they are independent of the metric which can be used to define the classical theory. \newline
There are two types of TFTs, the first are the so-called Witten-type models \cite{Witten:1988ze}, which main property is that the gauge-fixed action is a BRST-exact expression. A typical example for this type is the topological Yang-Mills model in four space-time dimensions. The other type are the Schwarz-type models, which representatives are Chern-Simons and BF theories. A common feature of topological models is the existence of the so-called topological vector supersymmetry. Its graded algebra with the BRST-operator is of Wess-Zumino type and therefore closes on space-time translations. \newline 
The ultraviolet and infrared perturbative finiteness of Schwarz-type models in the framework of algebraic renormalization has been widely discussed. However, recently the authors of \cite{DelCima:1999kb} proposed a new topological model in three space-time dimensions, which is an analog to the two dimensional model introduced by Chamseddine and Wyler \cite{Chamseddine}. It is obtained by a dimensional reduction from a BF model in $D=4$. In general this yields a BF model in $D=3$ with an additional metric independent term proportional to $\epsilon^\mnr K_\mn D_\rho\phi$. The model, which they called BFK model due to the occurrence of the field $K_\mn$, was proven to be finite to all orders of perturbation theory. \newline
In the present paper we couple the action of the BFK model to an additional Chern-Simons term. Our aim is to show that this model is perturbative finite. Due to the Chern-Simons coupling the model can not be inferred from a dimensional reduction any longer. This considerably influences the symmetry content of the theory. \newline
In \cite{Baulieu:1984ih} the author discussed a powerful formalism in order to quantize gauge-theories, which intimately relies on the considerations about a geometric interpretation of the BRST symmetry \cite{Thierry-Mieg:Bonora}. In \cite{Gieres:2000pv} the authors make use of this algebraic approach to obtain the vector supersymmetry transformations for Schwarz-type models as well as Witten-type theories. By enlarging this concept in order to involve anti-fields in the sense of Batalin and Vilkovisky \cite{Batalin:1983jr} this algorithm represents a very elegant method for the gauge-fixing procedure. In the case of a BF model this was already considered in \cite{Wallet:1990wr}. A brief introduction of new possible topological theories which are derived with the help of this method is given in \cite{Baulieu}. \newline
For the purpose of the algebraic renormalization procedure \cite{PiguetSorella:Schweda} we will use the concept of the BRST symmetry \cite{Becchi:1974md} and we will follow the track of \cite{PiguetSorella:Schweda,Delduc:1989ft} for Chern-Simons theory as well as for the BFK model \cite{DelCima:1999kb}. The central role in this framework will play the vector supersymmetry and a further scalar supersymmetry, denoted by ${\cal D}$-symmetry, which contrary to \cite{DelCima:1999kb} can not be obtained by dimensional reduction from four dimensions. \newline
The present work is organized as follows. Section 2 defines the classical action with its gauge symmetries. Section 3 is devoted to the gauge-fixing procedure in the spirit of the above mentioned formalism. Furthermore, we give the explicit BRST transformations, we construct both the vector supersymmetry as well as ${\cal D}$-symmetry transformations for all fields characterizing the model, and finally analyze the off-shell algebra. In section 4 we perform the proof of the finiteness of the theory by discussing the stability and the existence of possible anomalies at the quantum level with the help of cohomology techniques.

\section{The classical action and its symmetries}

\subsection{The classical action}

The classical action in three dimensional space-time of the model we consider is given by the Chern-Simons action plus the BFK-term. The BFK-action can either be thought of as a possible metric independent combination of a two-form, one-form and scalar field in three dimensions \cite{Baulieu} or as the action which stems from a dimensional reduction of a BF-model in $D=4$ \cite{DelCima:1999kb}. Our action looks like 
\bea
        S_{class}&=&\intM{3} \left\{\bh\left(AdA+\bzd AAA\right)+B_1F_2+K_2D\phi\right\},
\label{classaction}
\eea
where $A=A_\mu dx^\mu$ is the connection one-form with its corresponding curvature two-form $F_2=dA+AA$, $B_1=B_\mu dx^\mu$ is a one-form field, $K_2=\bh K_\mn dx^\mu dx^\nu$ is a two-form field and $\phi$ is a scalar. All fields take their values in the adjoint representation of some compact, semi-simple gauge group
\bea
        \varphi=\varphi^a T^a,
\eea
and the matrices $T^a$ are the generators of the Lie algebra, which are chosen to be anti-hermitian and obeying the relations
\bea
        [T^a,T^b]=f^{ab}{}_c T^c,&\mbox{and } \tr(T^a T^b)=\bh\delta^{ab}.
\eea
The covariant derivation $D$ on any field $\varphi$ is given by\footnote{The brackets are understood in a graded sense: $[A,B]=AB-(-1)^{|A||B|}BA$, where $|\Omega_p^q|=p+q$ defines the total grading of the form $\Omega_p^q$ which is given by the sum of the form-degree $|\Omega_p^q|_F=p$ and ghost-number $|\Omega_p^q|^{\Phi\Pi}=q$.}
\bea
        D\varphi=d\varphi+[A,\varphi].
\eea

\subsection{Gauge symmetries}

The classical action is invariant under the gauge-symmetry defined by
\be\ba{rclrcl}
        \delta \phi &=& [\lambda,\phi], 	& \delta B_1 &=& D\eta+[\lambda,B_1], \nm \\
        \delta A    &=& D\lambda,		& \delta K_2 &=& D\kappa_1+[\lambda,K_2], 
\label{gauge}
\ea\ee
where $\lambda$, $\eta$ and $\kappa_1$ are the Lie algebra valued gauge-parameters. The above gauge-transformations are reducible since the action is still invariant if we let $\kappa_1=D\kappa$.
 
\section{Batalin-Vilkovisky action and gauge-fixing}

\subsection{General setup}

One possible way of gauge-fixing the symmetries of the classical action (\ref{classaction}) is by turning the gauge-parameters $\lambda$, $\eta$ and $\kappa_1$ of (\ref{gauge}) into ghost-fields following the track of BRST quantization. However, we choose a slightly different approach to the subject which also leads to a BRST gauge-fixed action, but furthermore provides us with some nice features, such as a more transparent way of deriving the vector supersymmetry and the ${\cal D}$-symmetry. This approach rather follows the Batalin-Vilkovisky quantization procedure \cite{Batalin:1983jr} and is discussed in \cite{Gieres:2000pv}. \newline
We start by enlarging usual space-time by new coordinates in order to define generalized forms. The form-degree in the new directions is the ghost-number and a generalized form living in that space may be expanded in components
\bea
        \tilde X_p=X_d^{p-d}+X_{d-1}^{p-d+1}+\ldots+X_p+X_{p-1}^1+\ldots+X^p=\sum_{i=0}^d X_{d-i}^{p-d+i}.
\eea
where the lower index is the usual form-degree, the upper index labels the ghost-number and $d$ denotes the dimension of space-time. We can also define a so-called ``dual form'' in the spirit of \cite{Baulieu} which is given by
\bea
        \tilde Y_{d-p-1}=Y_d^{-p-1}+Y_{d-1}^{-p}+\ldots+Y_{d-p-1}+Y_{d-p-2}^1+\ldots+Y^{d-p-1}=\sum_{i=0}^d Y_{d-i}^{-p-1+i}.
\eea
The reason why these two forms are called dual to each other, is that the fields with negative ghost-charge serve as anti-fields in the sense of Batalin-Vilkovisky of the fields with positive ghost-number of the dual generalized form, i.e.
\bea
        X_{d-i}^{p-d+i}=(Y_i^{d-p-1-i})^*, \qquad 0\le i \le d-p-1, \qquad \mbox{and vice versa}.
\eea 
Thinking of this kind of superspace, we can generalize to an exterior derivative $\tilde d$ by
\bea
        \tilde d=d+s,
\eea
where $d$ is the ordinary exterior derivative and $s$ is the BRST-operator. The nilpotency of $\tilde d$ ensures $s^2=d^2=sd+ds=0$. Equipped with these ingredients we can build two pairs of dual forms in three dimensions
\be
\ba{rclrcl}
        \tilde A&=&A_3^{-2}+A_2^{-1}+A+c, 		& \tilde K_2&=&K_3^{-1}+K_2+K_1^1+K^2, \\
        \tilde B_1&=&B_3^{-2}+B_2^{-1}+B_1+B^1, 	& \tilde \phi&=&\phi_3^{-3}+\phi_2^{-2}+\phi_1^{-1}+\phi.
\label{genform}
\ea
\ee
The generalized connection $\tilde A$ admits to define a derivative $D^{\tilde A}=d+[\tilde A,.]$, whereas a covariant derivative is given by $\tilde D=\tilde d+[\tilde A,.]=s+D^{\tilde A}$.

\subsection{The minimal BV-action}

With these fields we are able to write down an action
\bea
        S_{gen}=\intM{3}\left.\left\{\bh\left(\tilde Ad\tilde A+\bzd \tilde A\tilde A\tilde A\right)+\tilde B_1D^{\tilde A}\tilde A+\tilde K_2D^{\tilde A}\tilde \phi\right\}\right|_3^0,
\label{genaction}
\eea
which transforms into a total derivative under the action of the BRST-operator $s$ following from the horizontality conditions
\be
\ba{rclrcl}
        \tilde d\tilde A+\bh[\tilde A,\tilde A]&=&0, & \tilde D\tilde K_2&=&0, \\
        \tilde D\tilde B_1&=&[\tilde K_2,\tilde \phi],& \tilde D\tilde \phi&=&0.
\label{hc}
\ea
\ee
By substitution of (\ref{genform}) into the action (\ref{genaction}) we get the classical action (\ref{classaction}), but furthermore terms where the fields with negative ghost charge are coupled to the BRST variations of the fields with positive ghost number plus additionally a three-linear term\footnote{This additional term is denoted by $S_{mod}$. In the four dimensional BF-model \cite{Guadagnini:1991br} it is implemented to restore the BRST-invariance of the gauge-fixed action. Since the BFK-model \cite{DelCima:1999kb} stems from a dimensional reduction this additional term exists there too. In the above procedure it is present automatically from the very beginning.}. Hence, if we identify all fields with negative ghost charge (or at least their linear combination) as anti-fields in the following way
\be
\ba{rclrcl}
        A_3^{-2}+B_3^{-2}&=&c^*,& K_3^{-1}&=&\phi^*, \nm \\
        A_3^{-2}&=&(B^1)^*,& \phi_3^{-3}&=&-(K^2)^*,\nm \\
        A_2^{-1}+B_2^{-1}&=&A^*,& \phi_2^{-2}&=&-(K_1^1)^*,\nm \\
        A_2^{-1}&=&(B_1)^*,&  \phi_1^{-1}&=&-(K_2)^*.
\ea
\ee
the action given in (\ref{genaction}) turns out to be the minimal action $S_{min}$ plus $S_{mod}$
\bea
        S_{gen}&=&S_{min}+S_{mod}\nm \\
        &=&S_{class}+\intM{3}\left\{ -c^*sc-A^*sA-(B^1)^*sB^1-(B_1)^*sB_1 \right. \nm \\
        && \left.-(K^2)^*sK^2-(K_1^1)^*sK_1^1-(K_2)^*sK_2-\phi^*s\phi\right\}+ S_{mod}.
\label{minaction}
\eea
$S_{mod}$ is given by
\bea
        S_{mod}=\intM{3}K^2[-(K_2)^*,(B_1)^*].
\label{modaction}
\eea
The anti-fields can by organized in $\Phi^*_a=(A^*,c^*,(B_1)^*,(B^1)^*,(K_2)^*,(K_1^1)^*,(K^2)^*,\phi^* )$, corresponding to the gauge-fields and ghosts $\Phi^a=(A,c,B_1,B^1,K_2,K_1^1,K^2,\phi)$. The BRST-transformations (\ref{hc}) clearly coincide with those obtained by the formula
\bea
        s\Phi^a=-\frac{\delta S_{min}}{\delta \Phi^*_a}.
\eea

\subsection{The BV-gauge-fixing procedure}

In a next step we can think about gauge-fixing which allows us to eliminate the anti-fields of the action (\ref{minaction}) but also from the generalized forms (\ref{genform}). In order to proceed that way we introduce the gauge-fermion\footnote{The ${}*$ denotes the Hodge-operator, in order to define a scalar product of forms $\langle \Omega_p,\Lambda_p\rangle=\intMohnetr{d}\Omega_p*\Lambda_p=\langle \Lambda_p,\Omega_p\rangle$. If the fields carry a $\Phi\Pi$-charge also the scalar product is given by $\langle \Omega_p^q,\Lambda_p^r \rangle=(-1)^{(d+p)(q+r)qr}\langle \Lambda_p^r,\Omega_p^q\rangle$.}
\bea
        \Psi_{gf}&=&\intM{3}\left\{\bar c_1^{-1}d*K_2+\bar c^{-2}d*K_1^1+\bar c^0(\alpha*\pi^{-1}+d*\bar c_1^{-1})+\bar cd*A+\bar \xi d*B_1\right\},
\eea
where $\alpha$ is an arbitrary gauge-parameter. With $\Psi_{fields}$ we fix the gauge-freedom for $A, B_1, K_2, K_1^1$ but also for $\bar c_1^{-1}$ which is present due to the reducible symmetry of $K_2$. 
The anti-ghosts and the corresponding multiplier fields are collected together in $\bar\Phi^{\alpha}_{anti}=(\bar c,\bar\xi,\bar c_1^{-1},\bar c^{-2},\bar c^0)$, $\Phi^{\alpha}_{mult}=(b,\lambda,\pi_1,\pi^{-1},\pi^1)$. The anti-ghosts come in trivial BRST-doublets $s\bar\Phi^{\alpha}_{anti}=\Phi^{\alpha}_{mult}$ and $s\Phi^{\alpha}_{mult}=0$ which is guaranteed by the additional action 
\bea
        S_{aux}=\intM{3}\left\{-\sum_{\alpha} (\bar\Phi^{\alpha}_{anti})^*\Phi^{\alpha}_{mult}\right\} =\intM{3}\left\{-(\bar c_1^{-1})^*\pi_1-(\bar c^{-2})^*\pi^{-1}-(\bar c^0)\pi^1-(\bar c)^*b-(\bar \xi)^*\lambda  \right\},
\eea
We also include external sources labelled by $\rho^*_a=(\gamma^*,\tau^*, \rho_2^{*-1},\rho_3^{*-2}, b_1^{*-1},b_2^{*-2},b_3^{*-3},\lambda_3^{*-1})$, coupled to the fields with non-linear BRST transformations $\Phi^a$. These sources are necessary in the further consideration to write down a Slavnov-Taylor operator. Henceforth, all gauge, ghost, anti-ghost and multiplier fields can be addressed by $\Phi^A=(\Phi^a,\bar\Phi_{anti}^{\alpha},\Phi_{mult}^{\alpha})$.\newline
The gauge-fermion for the external sources looks like
\bea
        \Psi_{ext}&=&\intM{3}\left\{\sum_a (-1)^{1+|\Phi^a|_F} \Phi^a \rho^*_a\right\} \\
        &=&\intM{3}\left\{-K_2 b_1^{*-1}+K_1^1 b_2^{*-2}-K^2 b_3^{*-3}+A\gamma^*-c\tau^*+B_1\rho_2^{*-1}-B^1\rho_3^{*-2}-\phi\lambda_3^{*-1}\right\}, \nm
\eea
which leads to the total gauge-fermion 
\bea
        \Psi=\Psi_{gf}+\Psi_{ext}.
\eea
The total action is now given by
\bea
        S=S_{gen}+S_{aux}=S_{min}+S_{mod}+S_{aux}.
\eea
The gauge-fermion serves to eliminate the anti-fields due to the formula
\be
        \Phi^*_A=-\frac{\delta \Psi}{\delta \Phi^A}.
\label{antiraus}
\ee
Finally, this elimination yields the total gauge fixed action 
\bea
        \g{0}&=&\left.\left[S_{min}+S_{mod}+S_{aux}\right]\right|_{\Phi^*_A=-\frac{\delta \Psi}{\delta \Phi^A}} \nm \\
        &=&S_{class}+S_{gf}+S_{ext}-K^2[b_1^{*-1}+*d\bar c_1^{-1},\rho_2^{*-1}+*d\bar\xi] 
\label{action}
\eea
where
\bea
        S_{gf}&=&s\Psi_{gf}, \\
        S_{ext}&=&s\Psi_{ext}=\intM{3}\left\{\sum_a \rho_a^*s\Phi^a\right\} \nm\\
        &=&\intM{3}\left\{b_1^{*-1}sK_2+b_2^{*-2}sK_1^1+b_3^{*-3}sK^2+\lambda_3^{*-1}s\phi+\gamma^*sA+\tau^*sc+\rho_2^{*-1}sB_1+\rho_3^{*-2}sB^1\right\}, 
\eea
The action (\ref{action}) may be rearranged to
\bea
        \g{0}&=&\g{0}_{CS}+\g{0}_{BF}+\g{0}_{KD\phi}-K^2[b_1^{*-1}+*d\bar c_1^{-1},\rho_2^{*-1}+*d\bar\xi], 
\eea
where the particular pieces are given by 
\bea
        \g{0}_{CS}&=&\intM{3}\left\{\bh\left(AdA+\bzd AAA\right)+bd*A+\left(\gamma^*+*d\bar c\right)sA+\tau^*sc\right\}, \\
        \g{0}_{BF}&=&\intM{3}\left\{B_1DA+\lambda d*B_1+\left(\rho_2^{*-1}+*d\bar\xi\right)sB_1+\rho_3^{*-2}sB^1\right\}, \\
        \g{0}_{KD\phi}&=&\intM{3}\left\{K_2D\phi+\pi_1d*K_2+\pi^{-1}d*K_1^1+\pi^1d*\bar c_1^{-1}+\alpha \pi^1*\pi^{-1}+\bar c^0d*\pi_1 \right. \nm \\
        && \left. +\left(b_1^{*-1}+*d\bar c_1^{-1}\right)sK_2+\left(b_2^{*-2}-*d\bar c^{-2}\right)sK_1^1+b_3^{*-3}sK^2+\lambda_3^{*-1}s\phi \right\}.
\eea

\subsection{Generalized forms}

By the elimination of the anti-fields via formula (\ref{antiraus}) the generalized forms (\ref{genform}) become 
\bea
        \tilde A&=& -\rho_3^{*-2}-\left(\rho_2^{*-1}+*d\bar\xi\right)+A+c, \nm \\
        \tilde B_1&=&-\left(\tau^*-\rho_3^{*-2}\right)-\left(\gamma^*-\rho_2^{*-1}+*d(\bar c-\bar\xi)\right)+B_1+B^1, \nm \\
        \tilde K_2&=&-\lambda_3^{*-1}+K_2+K_1^1+K^2, \nm \\
        \tilde \phi&=&b_3^{*-3}+\left(b_2^{*-2}-*d\bar c^{-2}\right)+\left(b_1^{*-1}+*d\bar c_1^{-1}\right)+\phi.
\eea
The components of the forms have the dimensions and $\Phi\Pi$-charges which are presented in table \ref{tab1}, \ref{tab2} and \ref{tab3}.
\begin{table}[ht]
\begin {center}
\begin{tabular}{|r|r|r|r|r|r|r|r|r|} \hline
        & $A$ & $c$ & $B_1$ & $B^1$ & $K_2$ & $K_1^1$ & $K^2$ & $\phi$\\ \hline
        dim & 1 & 0 & 1 & 0 & 2 & 1 & 0 & 0 \\ \hline
        $\Phi\Pi$ & 0 & 1 & 0 & 1 & 0 & 1 & 2 & 0 \\ \hline
\end{tabular}
\caption{Dimensions and Faddeev-Popov charges of $\Phi^a$}
\label{tab1}
\end{center}
\end{table}
\begin{table}[ht]
\begin {center}
\begin{tabular}{|r|r|r|r|r|r|r|r|r|r|r|} \hline
        & $\bar c$ & $b$ & $\bar\xi$ & $\lambda$ & $\bar c_1^{-1}$ & $\pi_1$ & $\bar c^{-2}$ & $\pi^{-1}$ & $\bar c^0$ & $\pi^1$ \\ \hline
        dim & 1  & 1 & 1 & 1 & 0 & 0 & 1 & 1 & 2 & 2\\ \hline
        $\Phi\Pi$ & -1 & 0 & -1 & 0 & -1 & 0 & -2 & -1 & 0 & 1\\ \hline
\end{tabular}
\caption{Dimensions and Faddeev-Popov charges of $\Phi_{anti}^\alpha$ and $\Phi_{mult}^\alpha$}
\label{tab2}
\end{center}
\end{table}
\begin{table}[ht]
\begin {center}
\begin{tabular}{|r|r|r|r|r|r|r|r|r|} \hline
        & $\gamma^*$ & $\tau^*$ & $\rho^{*-1}_2$ & $\rho_3^{*-2}$ & $b_1^{*-1}$ & $b_2^{*-2}$ & $b_3^{*-3}$ & $\lambda_3^{*-1}$\\ \hline
        dim & 2 & 3 & 2 & 3 & 1 & 2 & 3 & 3 \\ \hline
        $\Phi\Pi$ & -1 & -2 & -1 & -2 & -1 & -2 & -3 & -1 \\ \hline
\end{tabular}
\caption{Dimensions and Faddeev-Popov charges of $\rho^*_A$}
\label{tab3}
\end{center}
\end{table}

\subsection{BRST transformations and Slavnov-Taylor identity}

The action (\ref{action}) is invariant under the BRST transformations 
\be
\ba{c}
        sA=-Dc,\qquad sc=-c^2,\qquad s\phi=-[c,\phi], \nm \\
        sB_1=[K_1^1,\phi]+[K^2,*d\bar c_1^{-1}]-DB^1-[c,B_1], \nm \\
        sB^1=[K^2,\phi]-[c,B^1], \nm \\
        sK_2=-DK_1^1-[c,K_2]+[*d\bar\xi,K^2], \nm \\
        sK_1^1=-DK^2-[c,K_1^1], \qquad sK^2=-[c,K^2], \nm \\
\ea
\ee
\be
\ba{rclrcl}     
        s\bar c_1^{-1}&=&\pi_1,&s\pi_1&=&0, \nm \\
        s\bar c^{-2}&=&\pi^{-1},&s\pi^{-1}&=&0, \nm \\
        s\bar c^0&=&\pi^1,&s\pi^1&=&0, \nm \\
        s\bar c&=&b,&sb&=&0, \nm \\
        s\bar\xi&=&\lambda,&s\lambda&=&0.
\ea
\ee
The BRST invariance of $\g{0}$ is also expressed through the Slavnov-Taylor identity
\bea
        {\cal S}(\g{0})&=&\intM{3}\left\{\sum_a \ST{\rho^*_a}{\Phi^a}+\sum_{\alpha} \Phi_{mult}^{\alpha}\V{\g{0}}{\bar\Phi^{\alpha}_{anti}} \right\} \nm \\
        &=& \intM{3}\left\{\ST{\gamma^*}{A}+\ST{\tau^*}{c}+\ST{\rho_2^{*-1}}{B_1}+\ST{\rho_3^{*-2}}{B^1}+\ST{b_1^{*-1}}{K_2}+\ST{b_2^{*-2}}{K_1^1}\right.\nm\\
        &&\left.+\ST{b_3^{*-3}}{K^2}+\ST{\lambda_3^{*-1}}{\phi}+b\V{\g{0}}{\bar c}+\lambda\V{\g{0}}{\bar \xi}+\pi_1\V{\g{0}}{\bar c_1^{-1}}+\pi^{-1}\V{\g{0}}{\bar c^{-2}}+\pi^1\V{\g{0}}{\bar c^0}\right\}=0.
\label{ST}
\eea
For later purpose we introduce the linearized Slavnov-Taylor operator
\bea
        {\cal S}_{\g{0}}&=&\intM{3}\left\{\sum_a \left(\LST{\rho^*_a}{\Phi^a}\right)+\sum_{\alpha} \Phi_{mult}^{\alpha}\V{\g{0}}{\bar\Phi^{\alpha}_{anti}} \right\} \nm \\
        &=& \intM{3}\left\{\LST{\gamma^*}{A}+\LST{\tau^*}{c}+\LST{\rho_2^{*-1}}{B_1}\right. \nm \\
        & &\left.+\LST{\rho_3^{*-2}}{B^1}+\LST{b_1^{*-1}}{K_2}+\LST{b_2^{*-2}}{K_1^1}\right.\nm\\
        & &\left.+\LST{b_3^{*-3}}{K^2}+\LST{\lambda_3^{*-1}}{\phi}+b\V{\g{0}}{\bar c}+\lambda\V{\g{0}}{\bar \xi}+\pi_1\V{\g{0}}{\bar c_1^{-1}}\right. \nm \\
        & & \left. +\pi^{-1}\V{\g{0}}{\bar c^{-2}}+\pi^1\V{\g{0}}{\bar c^0}\right\}.
\label{LST}
\eea

\subsection{Vector supersymmetry}

On a flat space-time manifold the model under consideration exhibits an additional global invariance under the vector supersymmetry (for details see \cite{Gieres:2000pv}). The vector supersymmetry $\del=\tau^\mu\delta_\mu$\footnote{The constant parameter $\tau^\mu$ of the infinitesimal vector supersymmetry has ghost degree +2.} and the BRST-operator $s$ fulfill the on-shell algebra
\bea
        [s,\del]=\Li=[d,\id],
\label{algebra}
\eea
where $\Li$ is the Lie derivative along the constant vector $\tau^\mu$ and $\id$ the corresponding interior product. The algebra applied to the generalized forms (\ref{genform}) yields
\bea
        (\del s+s\del)\tilde \varphi-(\id d+d\id)\tilde \varphi&=&0,
\eea
where $\tilde\varphi=\{\tilde A,\tilde B_1,\tilde K_2,\tilde \phi\}$. With the help of the conditions (\ref{hc}) we can replace always the first and third term ($s\tilde \varphi$ and $d\tilde \varphi$). The definition $\tilde\id=\id-\del$ finally leads to the relations
\be
\ba{rclrcl}
        \tilde D\tilde\id\tilde A&=&0, & \tilde D\tilde\id\tilde K_2-[\tilde\id\tilde A,\tilde K_2]&=&0,  \\
        \tilde D\tilde\id\tilde B_1-[\tilde\id\tilde A,\tilde B_1]+[\tilde\id\tilde K_2,\tilde\phi]+[\tilde K_2,\tilde\id\tilde\phi]&=&0, & \tilde D\tilde\id\tilde \phi-[\tilde\id\tilde A,\tilde \phi]&=&0 .
\ea
\ee
Obviously, one possible solution is $\tilde\id \tilde\varphi=0$, hence, we have in a short-hand notation the $\del$-transformations
\be
\ba{rrrr}
        \del \tilde A=\id \tilde A, &\del \tilde B_1=\id \tilde B_1,&\del \tilde K_2=\id \tilde K_2,&\del \tilde \phi=\id \tilde \phi.
\ea
\ee
The algebra (\ref{algebra}) closes only modulo equations of motion
\be
\ba{rclrcl}
        \left[s,\del\right]A&=&\Li A-\id\V{\g{0}}{B_1}, 			& \left[s,\del\right]K_2&=&\Li K_2+\id\V{\g{0}}{\phi}, \\
        \left[s,\del\right]B_1&=&\Li B_1+\id\V{\g{0}}{A}-\id\V{\g{0}}{B_1}, 	& \left[s,\del\right]\phi&=&\Li \phi-\id\V{\g{0}}{K_2}.
\label{sddelalgebra}
\ea
\ee
On the remaining fields the algebra closes off-shell. If we choose $\alpha=-1$ the vector supersymmetry is indeed a symmetry of the action (\ref{action}), which is described by the Ward operator ${\cal W}_{(\tau)}=\tau^\mu{\cal W}_\mu$\footnote{$g(\tau)$ is defined as $\tau^\mu g_{\mu\nu} dx^\nu$. The Hodge-operator intertwines between the interior derivative $\id$ and the one-form $g(\tau)$ in the way $\id*\Omega_p^q=(-1)^p *g(\tau)\Omega_p^q$.}   
\bea
        {\cal W}_{(\tau)}&=&\intM{3}\left\{\id A\V{}{c}-\id\hat\rho_2^{*-1}\V{}{A}+\id B_1\V{}{B^1}-\id\left(\hat\gamma^*-\hat\rho_2^{*-1}\right)\V{}{B_1}+\id K_1^1\V{}{K^2}+\id K_2\V{}{K_1^1}  \right. \nm \\
        &&\left.-\id\lambda_3^{*-1}\V{}{K_2}+\id \hat b_1^{*-1}\V{}{\phi}+\Li\bar c\V{}{b}+\Li\bar\xi\V{}{\lambda}-g(\tau)\bar c^{-2}\V{}{\bar c_1^{-1}}+\left(\Li\bar c_1^{-1}-g(\tau)\pi^{-1}\right)\V{}{\pi_1}\right. \nm \\
        &&\left.+\Li \bar c^{-2}\V{}{\pi^{-1}}+\Li\bar c^0\V{}{\pi^1}+\id\tau^*\V{}{\gamma^*}+\id\rho_3^{*-2}\V{}{\rho_2^{*-1}}+\id b_2^{*-2}\V{}{b_1^{*-1}}+\id b_3^{*-3}\V{}{b_2^{*-2}}\right\},
\label{Wsusy}
\eea
where $\hat\rho_2^{*-1}, \hat\gamma^*, \hat b_1^{*-1}$ are given by 
\be
\ba{rclrcl}
        \hat\rho_2^{*-1}&=&\rho_2^{*-1}+*d\bar\xi, \\
        \hat\gamma^*&=&\gamma^*+*d\bar c, \\
        \hat b_1^{*-1}&=&b_1^{*-1}+*dc_1^{-1}, 
\label{hat}
\ea
\ee
However, the Ward identity is linearly broken in the quantum fields due to the external sources
\bea
        {\cal W}_{(\tau)} \g{0}=\Delta_{(\tau)},
\eea
where the linear breaking term is given by
\bea
        \Delta_{(\tau)}&=& \intM{3}\left\{ \sum_a (-1)^{|\Phi^a|}\Phi^*_a\Li\Phi^a+d\pi_1*\id\lambda_3^{*-1}+db*\id\rho_2^{*-1}+d\lambda*\id(\gamma^*-\rho_2^{*-1})\right\}.
\eea

\subsection{${\cal D}$-symmetry}

In \cite{DelCima:1999kb} the authors discussed the three-dimensional BFK-model in view of a dimensional reduction of a four-dimensional BF-model. Beside the vector supersymmetry the BFK-model is also invariant under a scalar supersymmetry with ghost-charge $-1$ which equals the fourth component of the vector supersymmetry of the BF-model. Surprisingly, the model under consideration also is invariant under a quite similar symmetry, which is denoted by ${\cal D}$-symmetry, although the model can not be reached by a dimensional reduction. In another shorthand notation the $\cal D$-transformations are given by
\be
\ba{rrrr}
        {\cal D}\tilde K_2=\tilde B_1,& {\cal D}\tilde A=-\tilde \phi,& {\cal D}\tilde B_1=\tilde \phi,& {\cal D} \tilde \phi=0.
\ea
\ee
The ${\cal D}$-transformations and the BRST-operator close on-shell 
\be
\ba{rclrcl}
        \left[s,{\cal D}\right]A&=&-\frac{\delta \g{0}}{\delta K_2},  & \left[s,{\cal D}\right]K_2&=&\frac{\delta \g{0}}{\delta A}-\frac{\delta \g{0}}{\delta B_1}, \\
        \left[s,{\cal D}\right]B_1&=&\frac{\delta \g{0}}{\delta K_2}, & \left[s,{\cal D}\right]\phi&=&0.
\ea
\ee
On the remaining fields $s$ and ${\cal D}$ anti-commutate off-shell. The symmetry is described via the Ward-identity
\bea
        {\cal W}^{\cal D}&=&\intM{3}\left\{-\phi\V{}{c}-\hat b_1^{*-1}\V{}{A}+\phi\V{}{B^1}+\hat b_1^{*-1}\V{}{B_1}+B^1\V{}{K^2}+B_1\V{}{K_1^1}-\left(\hat\gamma^*-\hat\rho_2^{*-1}\right)\V{}{K_2}\right.\nm\\
        &&\left.-\bar c^{-2}\V{}{\bar\xi}+\pi^{-1}\V{}{\lambda}+b_2^{*-2}\V{}{\rho_2^{*-1}}+b_3^{*-3}\V{}{\rho_3^{*-2}}+\left(\tau^*-\rho_3^{*-2}\right)\V{}{\lambda_3^{*-1}}      \right\},
\label{WD}
\eea
where $\hat b_1^{*-1}, \hat\gamma^*,\hat\rho_2^{*-1}$ are defined in (\ref{hat}).
Because of the external sources it is also linearly broken  
\bea
        {\cal W}^{\cal D}\g{0}&=&\Delta^{\cal D},
\eea
with
\bea
        \Delta^{\cal D}=\intM{3}\left\{b_1^{*-1}*d(b-\lambda)-(\gamma^*-\rho_2^{*-1})*d\pi_1\right\}.
\eea

\subsection{Gauge conditions, ghost and anti-ghost equations}

In order to prove the exact quantum scale invariance of the model under consideration we establish the gauge conditions, ghost and anti-ghost equations. The gauge conditions for $A$ and $B_1$ read as
\bea
        \frac{\delta \g{0}}{\delta b}=d*A,\quad\frac{\delta \g{0}}{\delta \lambda}=d*B_1,
\label{gaugecond1}
\eea
whereas the gauge conditions for $K_2,K_1^1,\bar c_1^{-1}$ and $\pi_1$ are
\bea
        \frac{\delta \g{0}}{\delta \pi_1}=d*K_2-*d\bar c^0, \quad\frac{\delta \g{0}}{\delta \pi^{-1}}=d*K_1^1+*\pi^1,\quad
        \frac{\delta \g{0}}{\delta \pi^1}=d*\bar c_1^{-1}-*\pi^{-1}, \quad\frac{\delta \g{0}}{\delta \bar c^0}= d*\pi_1.
\label{gaugecond2}
\eea
By commuting the Slavnov-Taylor identity with the gauge-conditions, on gets the following anti-ghost equations
\be
\ba{rclrcl}
        {\cal G}^A(\g{0})&=&\frac{\delta \g{0}}{\delta \bar c}+d*\frac{\delta \g{0}}{\delta \gamma^*}=0,&       {\cal G}^K_1(\g{0})&=&\frac{\delta \g{0}}{\delta \bar c_1^{-1}}-d*\frac{\delta \g{0}}{\delta b_1^{*-1}}=-*\pi^1,\nm \\
        {\cal G}^B(\g{0})&=&\frac{\delta \g{0}}{\delta \bar\xi}+d*\frac{\delta \g{0}}{\delta \rho_2^{*-1}}=0,& {\cal G}^K_2(\g{0})&=&\frac{\delta \g{0}}{\delta \bar c^{-2}}-d*\frac{\delta \g{0}}{\delta b_2^{*-2}}=0. 
\label{antighost}
\ea
\ee
The integrated ghost equations read
\bea
        \bar{\cal G}^B(\g{0})&=&\intMohnetr{3}\left\{\frac{\delta \g{0}}{\delta B^1}+\left[\bar\xi,\frac{\delta \g{0}}{\delta b}\right]\right\}=\bar\Delta^B, \nm \\
        \bar{\cal G}^K(\g{0})&=&\intMohnetr{3}\left\{\frac{\delta \g{0}}{\delta K^2}-\left[\bar c^{-2},\frac{\delta \g{0}}{\delta b}\right]\right\}=\bar\Delta^K,
\label{ghost}
\eea
where the linear breaking terms are given by
\bea
        \bar\Delta^B&=&\intMohnetr{3}\left\{\left[c,\rho_3^{*-2}\right]-\left[A,\rho_2^{*-1}\right]\right\}, \nm \\
        \bar\Delta^K&=&\intMohnetr{3}\left\{\left[b_1^{*-1},\rho_2^{*-1}+*d\bar\xi\right]-\left[b_2^{*-2},A\right]-\left[b_3^{*-3},c\right]-\left[\rho_2^{*-1},*d\bar c_1^{-1}\right]-\left[\rho_3^{*-2},\phi\right]\right\}.
\eea

\subsection{Off-shell algebra}

The following off-shell algebra is of major importance for the further considerations:
\be
\ba{rclrcl}
        {\cal S}_{\g{0}} {\cal S}_{\g{0}}&=&0, & \{ {\cal W}_{(\tau)},{\cal W}_{(\tau)} \}&=&0, \nm \\
        \{ {\cal W}^{\cal D},{\cal W}^{\cal D} \}&=&0, & \{ {\cal S}_{\g{0}},{\cal W}^{\cal D} \}&=&0, \nm \\
        \{ {\cal W}^{\cal D},{\cal W}_{(\tau)} \}&=&0,& \{  {\cal W}_{(\tau)},{\cal S}_{\g{0}} \}&=&{\cal P}_{(\tau)},
\ea
\ee
where 
\be
        {\cal P}_{(\tau)}=\intM{3} \sum_{A}\Li\Phi^A\V{}{\Phi^A}.
\label{WP}
\ee

\section{Proof of the finiteness}

This section is devoted to discuss the full symmetry content of the theory at the quantum level, i.e. the question of possible anomalies and the stability problem which amounts to analyze all invariant counterterms. 

\subsection{Stability}

In order to investigate the stability of the present model, we have to analyze the most general counterterms for the total action. This implies to consider the following perturbed action
\be 
        \Gamma=\Gamma^{(0)}+\Delta, 
\ee
where $\Gamma^{(0)}$ is the total action (\ref{action}) and $\Gamma$ is an arbitrary functional depending on the same fields as $\Gamma^{(0)}$ and satisfying the Slavnov-Taylor identity (\ref{ST}), the Ward identities for the vector supersymmetry (\ref{Wsusy}) and the ${\cal D}$-symmetry (\ref{WD}), the gauge conditions (\ref{gaugecond1}) and (\ref{gaugecond2}), the  anti-ghost equations (\ref{antighost}), the ghost equations (\ref{ghost}) and the Ward identity for the translations (\ref{WP}). The perturbation $\Delta$ collecting all appropriate invariant counterterms is an integrated local field polynomial of dimension three and ghost number zero. \newline
In a next step we take a closer look at the most general deformation of the classical action, which still has to fulfill the above constraints. In this spirit, the perturbation $\Delta$ has to obey the following set of equations:
\alpheqn
\parbox{4.7cm}{
\bea 
        {\delta\Delta\over\delta b} &=& 0 ,  \label{gc1} \\
        {\delta\Delta\over\delta\lambda} &=& 0 ,  \label{gc2} \\
        {\delta\Delta\over\delta\pi_1} &=& 0 ,  \label{gc3} \\
        {\delta\Delta\over\delta\pi^{-1}} &=& 0 ,  \label{gc4} \\
        {\delta\Delta\over\delta\pi^1} &=& 0 ,  \label{gc5} \\
        {\delta\Delta\over\delta\bar c^0} &=& 0 ,  \label{gc6} 
\eea}\hfill
\parbox{5.2cm}{
\bea
        {\cal S}_{\Sigma}\Delta &=& 0 ,  \label{con1} \\
        {\cal W}_\tau\Delta &=& 0 ,  \label{con2} \\
        {\cal W}^{\cal D}\Delta &=& 0 , \label{con3} \\ 
        {\cal P}_{(\varepsilon)}\Delta &=& 0 ,\label{con4} \\
        \intMohnetr{3}{\delta\Delta\over\delta B^1} &=& 0 , \label{con5} \\ 
        \intMohnetr{3}{\delta\Delta\over\delta K^2} &=& 0 . \label{con6} 
\eea}
\parbox{6.3cm}{
\bea
        {\delta\Delta\over\delta\bar c}+d*{\delta\Delta\over\delta\gamma^*} &=& 0 ,\label{a1}\\   
        {\delta\Delta\over\delta\bar\xi}+d*{\delta\Delta\over\delta\rho_2^{*-1}} &=& 0, \label{a2}\\
        {\delta\Delta\over\delta\bar c_1^{-1}}-d*{\delta\Delta\over\delta b_1^{*-1}} &=& 0 , \label{a3} \\
        {\delta\Delta\over\delta\bar c^{-2}}-d*{\delta\Delta\over\delta b_2^{*-2}} &=& 0 , \label{a4} 
\eea}\hfill
\reseteqn

One concludes from the first six equations (\ref{gc1})--(\ref{gc6}) that the perturbation $\Delta$ is independent of the multiplier fields $b$, $\lambda$, $\pi_1$, $\pi^{-1}$, $\pi^1$ and $\bar c^0$. The equations (\ref{a1})--(\ref{a4}) imply that dependence of $(\gamma^*,\bar c)$, $(\rho^{*-1}_2,\bar\xi)$, $(b^{*-1}_1,\bar c^{-1}_1)$ and $(b^{*-2}_2,\bar c^{-2})$ is given by $\hat\gamma^*, \hat\rho^{*-1}_2, \hat b^{*-1}_1$ defined in (\ref{hat}) and the following combination
\be 
        \hat b^{*-2}_2=b^{*-2}_2-*d\bar c^{-2}. 
\ee
The equations (\ref{con1})--(\ref{con4}), as in reference \cite{Becchi:1988vc}, can be unified into a single operator $\delta$:
\be 
        \delta={\cal S}_{\g{0}}+{\cal W}_{(\tau)}+\theta{\cal W}^{\cal D}+{\cal P}_{(\varepsilon)}+\intMohnetr{3}d^3x(-\tau^\mu){\de\over\de\varepsilon^\mu}+\intMohnetr{3}d^3x(-\theta){\de\over\de\eta}, 
\label{operatordelta} 
\ee
producing a cohomology problem 
\be 
        \delta\Delta=0 . 
\label{cohproblem} 
\ee
The constant vector $\varepsilon^\mu$ has ghost charge +1, whereas $\theta$ and $\eta$ are constant scalars carrying ghost number +2 and +1 respectively. It can be easily verified that the operator $\delta$ is nilpotent
\be 
        \delta^2=0 . 
\ee
Therefore, any expression  of the form $\delta\hat\Delta$ automatically satisfies (\ref{cohproblem}). A solution of this type is called a trivial solution. Hence, the most general solution of (\ref{cohproblem}) reads 
\be 
        \Delta=\Delta_c+\delta\hat\Delta . 
\ee
The nontrivial solution $\Delta_c$ is $\delta$-closed ($\delta\Delta_c=0$), however it is not $\delta$-exact ($\Delta_c\ne\delta\hat\Delta$). \newline
For the determination of the nontrivial solution of (\ref{cohproblem}), we need to introduce a filtering operator ${\cal N}$:
\be 
        {\cal N}=\intM{3}\sum_{\varphi}\varphi{\delta\over\delta\varphi} \qquad, 
\ee
where $\varphi$ stands for all fields, including $\tau, \varepsilon$, $\theta$ and $\eta$. To all fields we assign the homogeneity degree 1. The filtering operator induces a decomposition of $\delta$ according to
\be 
        \delta=\delta_0+\delta_1 . 
\label{decomp}
\ee
The operator $\delta_0$ does not increase the homogeneity degree while acting on a field polynomial. On the other hand, the operator $\delta_1$ increases the homogeneity degree by one unit. Furthermore, the nilpotency of $\delta$ leads now to 
\bea 
        \delta_0^2=0, & \{\delta_0,\delta_1\}=0,& \delta_1^2=0 . 
\label{relat} 
\eea
Hence, we obtain from (\ref{relat}) the following relation
\be 
        \delta_0\Delta=0, 
\ee
which yields a further cohomology problem. The usefulness of the decomposition (\ref{decomp}) relies on a very general theorem \cite{PiguetSorella:Schweda,Dixon:1991wi} stating that the cohomology of the complete operator $\delta$ is isomorphic to a subspace of the cohomology of the operator $\delta_0$, which is easier to solve than the cohomology of $\delta$. The operator $\delta_0$ acts on the fields as follows:
\be\ba{rclrclrclrcl}
        \delta_0 A&=&-dc,       & \delta_0 K_2&=&-dK_1^1,               & \delta_0\hat\rho^{*-1}_2&=&dA,                & \delta_0\lambda^{*-1}_3&=&-dK_2,   \\
        \delta_0 c&=&0,         & \delta_0 K^1_1&=&-dK^2,               & \delta_0\hat\rho^{*-2}_3&=&-d\hat\rho^{*-1}_2,& \delta_0 \varepsilon^{\mu}&=&-\tau^\mu , \\
        \delta_0 B_1&=&-dB^1,   & \delta_0 K^2&=&0,                     & \delta_0\hat b^{*-1}_1&=&d\phi ,              & \delta_0\tau^\mu&=&0, \\
        \delta_0 B^1&=&0,       & \delta_0\hat\gamma^*&=&dA+dB_1,       & \delta_0\hat b^{*-2}_2&=&d\hat b^{*-1}_1 ,    & \delta_0\eta&=&-\theta, \\
        \delta_0 \phi&=&0,      & \delta_0\tau^*&=&-d\hat\gamma^* ,     & \delta_0\hat b^{*-3}_3&=&d\hat b^{*-2}_2,     & \delta_0 \theta&=&0.
\ea\ee
We notice that the quantities $\varepsilon^\mu$, $\tau^\mu$ and $\eta$, $\theta$ respectively transform under $\delta_0$ as doublets, being therefore out of the cohomology \cite{Brandt}. The nontrivial solution $\Delta_c$ can now be written as integrated local field polynomial of form degree three and ghost number zero:
\be 
        \Delta_c=\intMohnetr{3}\omega^0_3 , 
\ee
where $\omega^p_q$ is a field polynomial of form degree $q$ and ghost number $p$. Using the Stoke's theorem, the Poincar\'e lemma \cite{Brandt} and the relation  $\{\delta_0,d\}=0$, we obtain the following tower of descent equations:
\be\ba{rclrcl}
        \delta_0\omega_3^0+d\omega^1_2 &=& 0 , & \delta_0\omega_1^2+d\omega^3_0 &=& 0 ,\\
        \delta_0\omega_2^1+d\omega^2_1 &=& 0 , & \delta_0\omega^3_0 &=&0 . 
\label{tower}
\ea\ee
In order to solve the tower of descent equations (\ref{tower}) we follow the technique of \cite{PiguetSorella:Schweda,Sorella:1992dr} and decompose the exterior derivative according to 
\bea
        [\bar\delta,\delta_0]=d, & [\bar\delta,d]=0, 
\eea
where the operator $\bar\delta$ is given by
\be
\ba{rclrclrclrcl}
	\bar\delta A&=&-2\hat\rho^{*-1}_2,      		& \bar\delta K_2&=&3\lambda^{*-1}_3,	& \bar\delta\hat\gamma^*&=&3\tau^*,                     	& \bar\delta\hat b^{*-1}_1&=&-2\hat b^{*-2}_2,    \\
	\bar\delta c&=&A,                       		& \bar\delta K^1_1&=&2K_2 ,		& \bar\delta\tau^*&=&0,                                 	& \bar\delta\hat b^{*-2}_2&=&-3\hat b^{*-3}_3 ,  \\
	\bar\delta B_1&=&-2\hat\gamma^*+2\hat\rho^{*-1}_2,	& \bar\delta K^2&=&K_1^1 ,		& \bar\delta\hat\rho^{*-1}_2&=&3\hat\rho^{*-2}_3,	& \bar\delta\hat b^{*-3}_3&=&0,   \\
	\bar\delta B^1&=&B_1,                                 	& \bar\delta \phi&=&-\hat b^{*-1}_1,    & \bar\delta\hat\rho^{*-2}_3&=&0,				& \bar\delta\lambda^{*-1}_3&=&0.
\label{delta-}
\ea
\ee
The benefit of the operator $\bar\delta$ is that $\omega^0_3$ is simply given by
\be 
        \omega^0_3=\bar\delta\bar\delta\bar\delta\;\omega_0^3 . 
\ee

The most general form for $\omega_0^3$ is constrained by the ghost number and form degree. Due to the fact that the field $\phi$ carries both vanishing ghost number and vanishing form degree it can appear an infinite number of times\footnote{For a model containing two scalar fields with vanishing $\Phi\Pi$-charge see \cite{Leitgeb}.} in $\omega_0^3$. Therefore, the latter reads
\bea 
        \omega_0^3      &=&     \sum_{i,j,k=0}^\infty\alpha_{ijk}\tr\left[c\phi^{i}c\phi^{j}c\phi^{k}\right]+ \sum_{i,j=0}^\infty\beta_{ij}\tr\left[c\phi^{i}K^2\phi^{j}\right]+\sum_{i,j,k=0}^\infty\gamma_{ijk}\tr\left[c\phi^{i}c\phi^{j}B^1\phi^{k}\right]+        \nm \\   
                        &+&     \sum_{i,j,k=0}^\infty\bar\alpha_{ijk}\tr\left[c\phi^{i}B^1\phi^{j}B^1\phi^{k}\right]+\sum_{i,j,k=0}^\infty\bar\beta_{ijk}\tr\left[B^1\phi^{i}B^1\phi^{j}B^1\phi^{k}\right]+\sum_{i,j=0}^\infty\bar\gamma_{ij}\tr\left[B^1\phi^{i}K^2\phi^{j}\right]. 
\label{omega03}
\eea
Here, the quantities $\alpha_{ijk},\beta_{ij},\gamma_{ijk},\bar\alpha_{ijk},\bar\beta_{ijk}$ and $\bar\gamma_{ij}$ stand for constant and field independent coefficients, which have to be determined. The upper indices of the field $\phi$ are just integer exponents required by locality. With the help of the operator $\bar\delta$ given in (\ref{delta-}) one can now easily calculate $\omega_3^0$. A careful and lengthy investigation shows that each monomial of (\ref{omega03}) leads to an expression which is forbidden by the ghost equations (\ref{con5}) and (\ref{con6}). Therefore, all of the coefficients $\alpha_{ijk},\beta_{ij},\gamma_{ijk},\bar\alpha_{ijk},\bar\beta_{ijk}$ and $\bar\gamma_{ij}$ in (\ref{omega03}) must be equal to zero. Consequently, we deduce that nontrivial solutions of both the $\delta_0$ cohomology as well as $\delta$ cohomology are empty. \newline
The calculation of the trivial solution of (\ref{cohproblem}) is straightforward. One has to find all possible counterterms of $\delta\hat\Delta$, which fulfill ghost number and form degree requirements. In fact, $\hat\Delta$ is a local field monomial of ghost number $-1$ and form degree 3. Again, since the field $\phi$ has both ghost number and form degree zero, it can appear an infinite number of times in the counterterms. Moreover, the expression $\delta\hat\Delta$ may depend also on the parameters $\varepsilon^\mu,\tau^\mu,\theta$ and $\eta$ which do not appear in the total action (\ref{action}). That is why the trivial counterterms must be independent of them. In other words, $\hat\Delta$ must be invariant under the vector supersymmetry, translations and $\cal D$-symmetry. A detailed and tedious analysis of this situation shows that there do not exist any possible field monomials for $\hat\Delta$ which obey the above conditions. Therefore, one concludes that the trivial counterterm vanishes identically.
 
\subsection{Search for anomalies}

The last problem to overcome in the proof of finiteness is the anomaly analysis. In the framework of renormalization theory one has to investigate whether the symmetries collected in $\delta$ are disturbed by quantum corrections. According to the quantum action principle, the symmetry breaking is described by
\be 
        \delta\Gamma={\cal A}, 
\ee
where ${\cal A}$ is a local, integrated, Lorentz-invariant field polynomial of form degree 3 and ghost number 1, that fulfills 
\be 
        \delta{\cal A}=0.
\label{anomalycoh} 
\ee
Due to the nilpotency of $\delta$ this defines a further cohomology problem. Writing ${\cal A}=\intMohnetr{3}\omega_3^1$ we are able to derive the following tower of descent equations by using the same strategy as in the previous section: 
\be\ba{rclrcl}
        \delta_0\omega_3^1+d\omega^2_2 &=& 0, & \delta_0\omega_1^3+d\omega^4_0 &=& 0, \\
        \delta_0\omega_2^2+d\omega^3_1 &=& 0, & \delta_0\omega^4_0 &=&0.
\label{anomalytower} 
\ea\ee  
The most general solution of last equation of (\ref{anomalytower}) is again constrained by the ghost number, form degree and the fact that the scalar $\phi$ can appear an infinite number of times in $\omega_0^4$. For $\omega_0^4$ we obtain:
\bea
        \omega_0^4      &=& \sum_{i,j,k,l=0}^\infty\alpha_{ijkl}\tr\left[c\phi^ic\phi^jc\phi^kc\phi^l\right]+\sum_{i,j,k=0}^\infty\beta_{ijk}\tr\left[c\phi^ic\phi^jK^2\phi^k\right]+\sum_{i,j=0}^\infty\gamma_{ij}\tr\left[K^2\phi^iK^2\phi^j\right] \nm \\
                        &+& \sum_{i,j,k,l=0}^\infty\delta_{ijkl}\tr\left[c\phi^ic\phi^jc\phi^kB^1\phi^l\right]+\sum_{i,j,k,l=0}^\infty\tau_{ijkl}\tr\left[c\phi^ic\phi^jB^1\phi^kB^1\phi^l\right]+\nm \\
                        &+& \sum_{i,j,k,l=0}^\infty\sigma_{ijkl}\tr\left[c\phi^iB^1\phi^jB^1\phi^kB^1\phi^l\right]+\sum_{i,j,k,l=0}^\infty\bar\alpha_{ijkl}\tr\left[B^1\phi^iB^1\phi^jB^1\phi^kB^1\phi^l\right]+\nm\\
                        &+& \sum_{i,j,k=0}^\infty\bar\beta_{ijk}\tr\left[c\phi^iB^1\phi^jK^2\phi^k\right]+\sum_{i,j,k=0}^\infty\bar\gamma_{ijk}\tr\left[B^1\phi^iB^1\phi^jK^2\phi^k\right]+\nm\\
                        &+& \sum_{i,j,k,l=0}^\infty\bar\delta_{ijkl}\tr\left[c\phi^iB^1\phi^jc\phi^kB^1\phi^l\right]+\sum_{i,j,k=0}^\infty\bar\tau_{ijk}\tr\left[c\phi^iK^2\phi^jB^1\phi^k\right].
\label{omega04}
\eea
Here, the quantities $\alpha_{ijkl},\beta_{ijk},\gamma_{ij},\delta_{ijkl},\tau_{ijkl},\sigma_{ijkl},\bar\alpha_{ijkl},\bar\beta_{ijk},\bar\gamma_{ijk}, \bar\delta_{ijkl}$ and $\bar\tau_{ijk}$ are constant field independent coefficients. Using the decomposition operator (\ref{delta-}), the solution to the descent equations reads
\be
        \omega^1_3=\bar\delta\bar\delta\bar\delta\omega_0^4.
\ee
Following the same arguments as in the previous section one can prove that all of the constant coefficients in (\ref{omega04}) must vanish. Therefore, the most general solution of $\delta{\cal A}=0$ is a $\delta$-exact quantity given by ${\cal A}=\delta\hat{\cal A}$ implying that the Slavnov identity, the Ward identities for the vector supersymmetry and the $\cal D$-symmetry as well as translations are anomaly free and can be promoted to the quantum level. Furthermore, following \cite{PiguetSorella:Schweda} one can easily show that the gauge conditions (\ref{gaugecond1}) and (\ref{gaugecond2}) as well as the anti-ghost equations (\ref{antighost}) are valid at the quantum level. Concerning the ghost equations (\ref{ghost}), it can also be proven to hold at the quantum level \cite{Blasi:1991xz}. \newline
As conclusion, the model we discussed is anomaly free and finite to all orders of perturbation theory.

\section{Acknowledgements}

The authors would like to thank J.Grimstrup and M.Schweda for fruitful discussions about the BFK model and the subject of cohomology.


\providecommand{\href}[2]{#2}\begingroup\raggedright\endgroup

\end{document}